\documentclass{article}
\usepackage{spconf,amsmath,graphicx,color,soul}
\usepackage{url}

\usepackage{lipsum} 

\usepackage{musicography}
\usepackage{enumitem}
\usepackage{siunitx}
\usepackage{hyperref}
\usepackage{cleveref}
\usepackage{footmisc}
\usepackage{booktabs}
\usepackage{cite}
\usepackage[utf8]{inputenc}
\DeclareUnicodeCharacter{266D}{\flat}
\DeclareUnicodeCharacter{266E}{\natural}
\DeclareUnicodeCharacter{266F}{\sharp}

\newcommand{\fref}[1]{Figure~\ref{#1}}
\newcommand{\tref}[1]{Table~\ref{#1}}

\title{Context-Aware Prosody Correction for Text-Based Speech Editing}
\name{Max Morrison$^{\flat,*}$ Lucas Rencker$^{\sharp,*}$\thanks{$^*$This work was carried out during an internship at Adobe Research.}, Zeyu Jin$^{\natural}$, Nicholas J. Bryan$^{\natural}$, Juan-Pablo Caceres$^{\natural}$, Bryan Pardo$^{\flat}$}
\address{$^{\flat}$Northwestern University, Evanston, IL, USA\\
         $^{\sharp}$University of Surrey, Guildford, UK\\
$^{\natural}$Adobe Research, San Francisco, CA, USA}

\begin{document}
\maketitle

\begin{abstract}
Text-based speech editors expedite the process of editing speech recordings by permitting editing via intuitive cut, copy, and paste operations on a speech transcript. A major drawback of current systems, however, is that edited recordings often sound unnatural because of prosody mismatches around edited regions. In our work, we propose a new context-aware method for more natural sounding text-based editing of speech. To do so, we 1) use a series of neural networks to generate salient prosody features that are dependent on the prosody of speech surrounding the edit and amenable to fine-grained user control 2) use the generated features to control a standard pitch-shift and time-stretch method and 3) apply a denoising neural network to remove artifacts induced by the signal manipulation to yield a high-fidelity result. We evaluate our approach using a subjective listening test, provide a detailed comparative analysis, and conclude several interesting insights.
\end{abstract}

\begin{keywords}
speech, prosody generation, pitch-shifting, time-stretching, deep learning
\end{keywords}


\section{Introduction}
\label{sec:intro}


Editing speech recordings based on text transcriptions is an emerging audio editing paradigm~\cite{Rubin_2013, jin2017voco, descriptref}. Given a time-aligned text transcription, content creators can quickly edit a transcript using familiar word processing operations such as cut, copy, and paste and automatically propagate changes to the corresponding audio recording without having to manually edit a raw waveform. However, when cutting or copying-and-pasting a word from one location to another in a recording, a mismatch in \emph{prosody}~\cite{nooteboom1997prosody} can occur between the inserted word and its context (e.g. mismatches in intonation, stress, or rhythm).


\begin{figure}[t]
	\centering
	\includegraphics[width=0.45\textwidth]{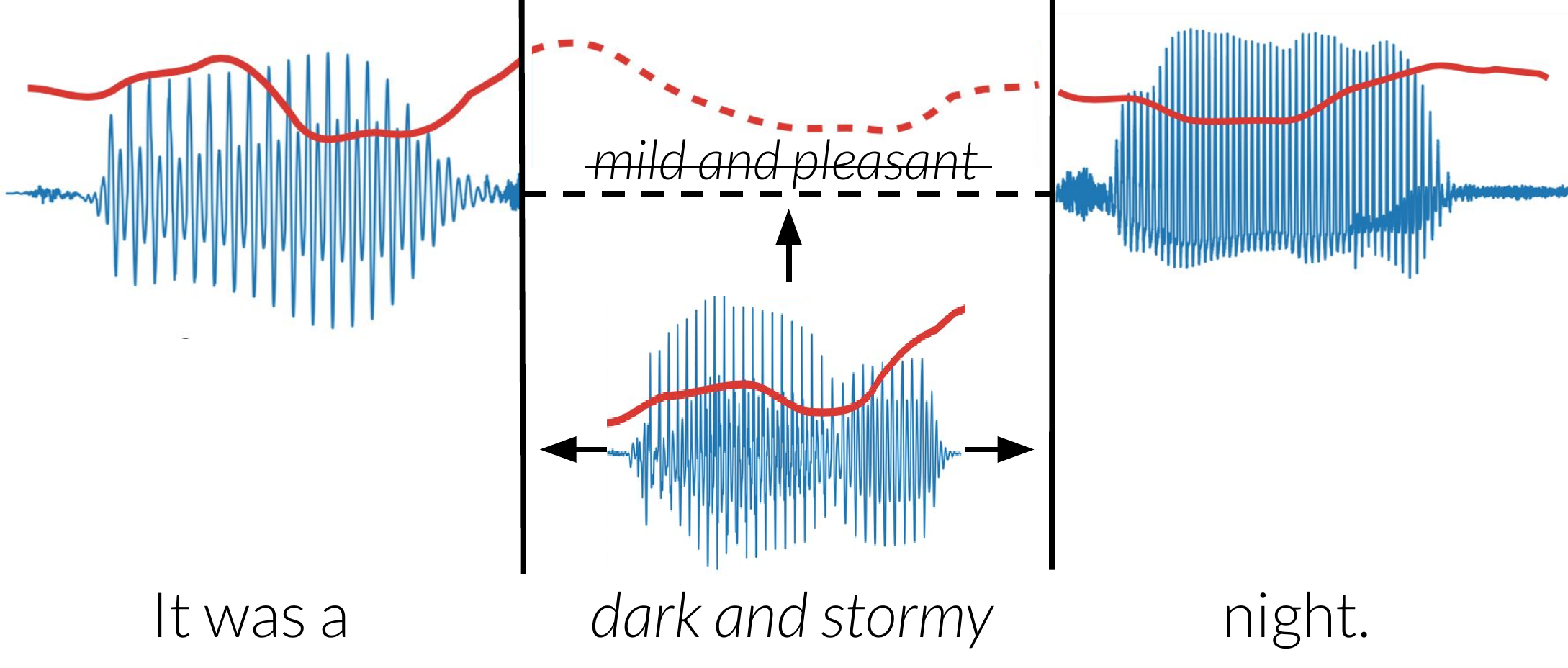}
	\caption{Context-aware text-based speech editing. We allow speech recordings (blue) to be manipulated by text-based cut, copy, and paste operations and then perform prosody correction (red) to make the edit sound natural in context.}
	\label{fig:speech_editing}
\end{figure}

\begin{figure}[t]
    \centering
    \includegraphics[width=\linewidth]{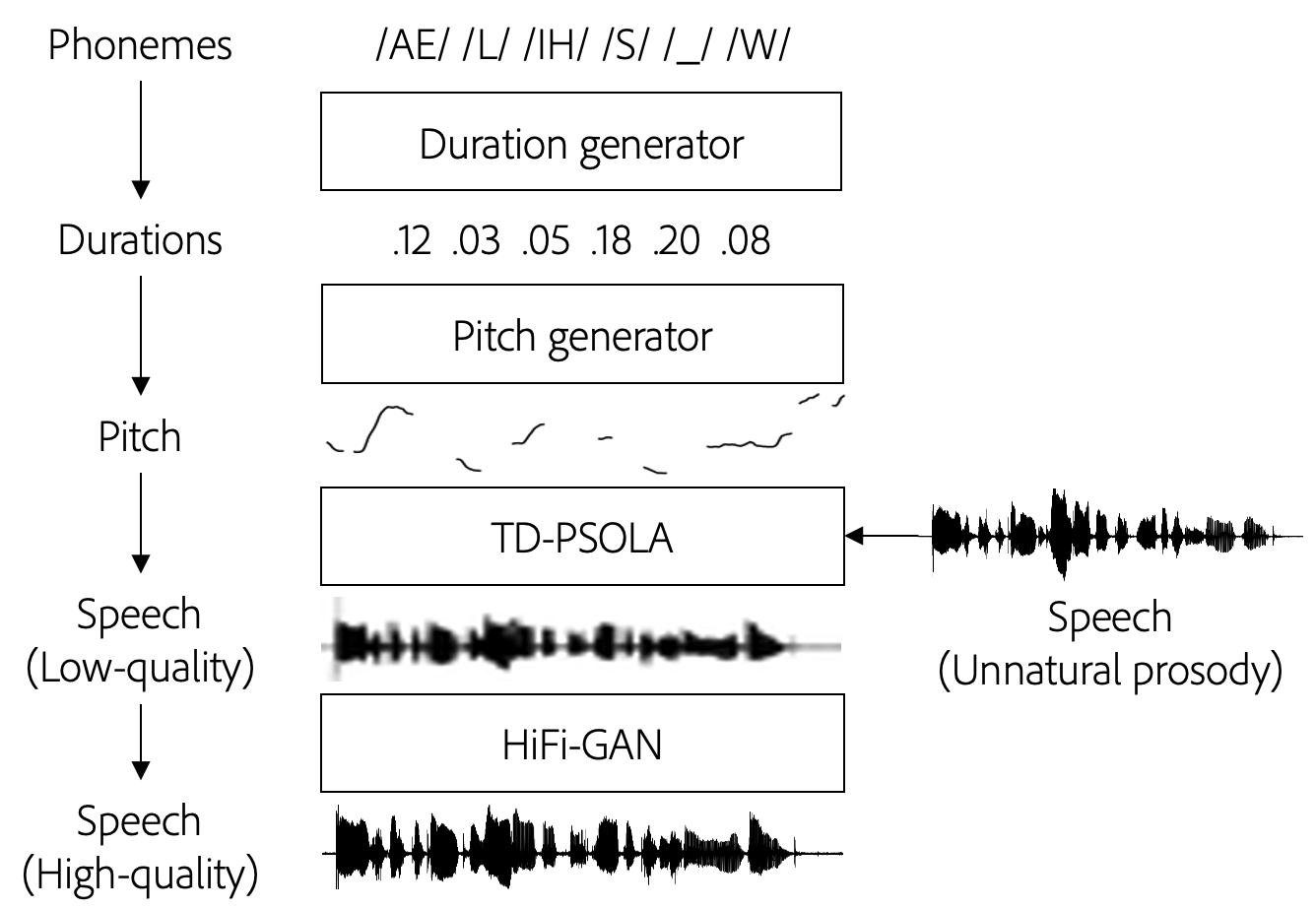}
    \caption{An overview of our proposed system. We first generate new, context-aware prosody from input phonemes using our pitch and duration generators. Next, we impose the generated prosody on the speech by pitch-shifting and time-stretching via TD-PSOLA. Finally, we use HiFi-GAN to mitigate the artifacts caused by pitch-shifting and time-stretching.}
    \label{fig:models}
\end{figure}

Prior work relevant to this problem includes neural prosody generation, speech manipulation (e.g., pitch-shifting and time-stretching), and speech editing. Previous work in neural prosody generation includes generating phoneme durations from text~\cite{ren2019fastspeech, ren2020fastspeech} and generating pitch from linguistic features and phoneme durations~\cite{wang2018fundamental, ren2020fastspeech}. Recently, Fastspeech 2~\cite{ren2020fastspeech} was the first neural network to explicitly generate both pitch and duration from text. However, these prosody generators cannot be independently trained and require a complex training setup involving spectrogram supervision and acoustic feature generation. More critically, FastSpeech 2 does not attempt to generate prosodies that sound natural in context with any preceding or following speech.


Relevant speech manipulation work includes fast pitch-shifting and time-stretching techniques suitable for real-time interactive applications. Digital signal processing (DSP) based methods for pitch-shifting and time-stretching speech include Time-Domain Pitch-Synchronous Overlap-and-Add (TD-PSOLA)~\cite{moulines1990pitch}, WORLD~\cite{Morise_2016}, and STRAIGHT~\cite{Moulines_1990}. These methods are computationally cheap and amenable to interactive applications, but can induce audible artifacts.  Real-time neural vocoders such as WaveGlow~\cite{prenger2019waveglow}, MelGAN~\cite{kumar2019melgan}, and LPCNet~\cite{valin2019lpcnet} tend to exhibit higher perceptual quality than DSP-based methods, but lack well-known pitch-shifting capabilities. More recently, FastPitch~\cite{lancucki2020fastpitch}  has demonstrated real-time, high-quality pitch-shifting and time-stretching, but requires significant retraining to add a new speaker and lacks context-aware generation. Finally, text-based speech editing methods such as VoCo~\cite{jin2017voco} and Descript~\cite{descriptref} are capable of performing cut, copy, and paste operations, but exhibit flattened prosody or boundary artifacts (i.e., prosody discontinuities at edit region boundaries).


In this paper, we propose a new system that improves text-based speech editing by allowing natural-sounding cut, copy, and paste operations. We perform the cut, copy or paste operation on the specified text and corresponding audio using a phoneme alignment transcription and then correct unnatural prosody regions using a series of four components as shown in \fref{fig:models}. The transformation consists of (1) a neural network that generates phoneme durations (including silence) from text, (2) a neural network that generates pitch from phonemes and corresponding durations, (3) pitch-shifting and time-stretching via TD-PSOLA~\cite{Moulines_1990} to impose the generated prosody (i.e., duration and pitch), and (4) a denoising and dereverberation neural network that removes manipulation artifacts of the edited speech~\cite{su2020hifi}. Our pitch and duration generators are amenable to fine- or coarse-grained user control, and can generate natural prosody in context with any preceding and following speech. We evaluate our approach using a subjective listening test, show a detailed comparative analysis, and conclude several interesting insights\footnote{Audio examples are available at \url{https://maxrmorrison.com/sites/context-aware}.}

\section{Methods}

\subsection{Phoneme duration generation}
\label{sec:durat}

We generate phoneme duration information (i.e., a value in seconds for each phoneme, including silence) for all words of an edit region using the first neural network in our system. To do so, we  break each word into phonemes and generate a duration value in seconds for each phoneme in the input text. Input phonemes are extracted using \texttt{g2pE}~\cite{g2pE2019} and one-hot encoded. The output of our neural network is a real-valued scalar in seconds that is upper-bounded at 0.5 seconds to prevent long silences.

Our duration generator is a sequence-to-sequence model consisting of an encoder and decoder. The encoder contains two 1D convolution blocks followed by a bidirectional gated recurrent unit (GRU). Each convolution block consists of a convolution with 512 channels and a kernel size of 5, a ReLU activation, and batch normalization~\cite{ioffe2015batch}. The decoder is a unidirectional GRU with 256 channels followed by a linear layer. During training, the network minimizes the mean squared error between the real-valued phoneme durations predicted by the network and the ground truth durations in seconds extracted with the Penn Phonetic Forced Aligner~\cite{yuan2008speaker}. We also experimented with using an autoregressive sequence-to-sequence model expecting the learned distribution of phoneme durations would be multimodal, but found that the distribution was unimodal and  hypothesize this is due to training on single-speaker data with a consistent reading style.

\subsection{Pitch generation}
Given the duration of each phoneme in a sentence being edited, we use a second neural network to generate a pitch contour for the edited words. More specifically, we use the existing Controllable DAR (C-DAR)~\cite{morrison2020controllable} model for pitch generation. C-DAR generates a pitch value for each 10 millisecond frame of speech from one-hot encoded phonemes and linguistic features (see \cite{morrison2020controllable}) that have been upsampled according to input phoneme durations. C-DAR predicts a categorical distribution over 128 possible pitch values, which are evenly distributed between -4 and +4 standard deviations from the speaker's average pitch in base-2 log-space. Ground truth pitch values are extracted using a PyTorch port~\cite{morrison2020torchcrepe} of the Crepe pitch tracker~\cite{kim2018crepe}. To minimize double and half frequency errors, we decode the pitch from the sequence of categorical distributions predicted by Crepe using Viterbi decoding. Voiced/unvoiced tokens are extracted by performing hysteresis thresholding on Crepe's network confidence value.

\subsection{Pitch-shifting and time-stretching}
To apply the generated prosody to edited speech, we use Time-Domain Pitch-Synchronous Overlap-and-Add (TD-PSOLA)~\cite{moulines1990pitch}. We choose TD-PSOLA over the WORLD~\cite{Morise_2016} vocoder due to its superior naturalness and pitch-shifting~\cite{morrison2020controllable}. TD-PSOLA works by pitch-tracking the speech to determine locations (\textit{pitch epochs}) to split the signal into chunks corresponding to one period. These chunks can then be repeated to perform time-stretching, or shifted to overlap more or less to adjacent chunks to perform pitch-shifting. The modified chunks are combined via Overlap-Add to create the pitch-shifted and time-stretched speech signal. We use the Praat~\cite{boersma2006praat} implementation of TD-PSOLA via the Parselmouth~\cite{jadoul2018introducing} Python package. We provide values for the pitch and time-stretching rate for each 10 millisecond frame of speech being edited. The time-stretching rate is the sequence of ratios between the original and generated phoneme durations, each repeated by the number of frames occupied by the original phoneme.

\subsection{Speech manipulation artifact removal}
\label{sec:hifi}
TD-PSOLA induces some artifacts in the edited speech. These artifacts are more noticeable in frames where duration or pitch change is larger. To mitgate this issue, we use HiFi-GAN~\cite{su2020hifi}, a GAN-based neural network that learns to transform noisy and reverberant speech recordings to clean, high-fidelity speech. While HiFi-GAN is only trained to perform speech denoising and dereverberation, we find it very effective at reducing the artifacts induced by pitch-shifting and time-stretching. This allows us to avoid using complex, speaker-dependent neural vocoders that can be difficult to use for interactive applications. HiFi-GAN also operates significantly faster than real-time (i.e. about 20x real-time on a Tesla V100 GPU), making it well-suited for interactive applications. More broadly, we believe the use of neural denoisers to mitigate DSP-based manipulation algorithm artifacts is an exciting research direction.

\begin{table*}
\centering
\begin{tabular}{r|cccccccccc}
 \textbf{} & \textbf{Original} & \textbf{Naive} & \textbf{Average} & \textbf{Proposed} & \textbf{-Duration} & \textbf{-Pitch} & \textbf{-HiFi-GAN} & \textbf{-Context} & \textbf{Tacotron}\\
\hline
\textbf{MOS} & 4.56 & 2.87 & 2.76 & \textbf{3.00} & 2.97 & 2.86 & 2.87 & 2.59 & 2.89 \\
\textbf{Pairwise} & 8.56\% & 57.7\% & 62.7\% & - & 45.3\% & 55.8\% & 54.2\% & 67.9\% & 53.2\%\\
\end{tabular}
\caption{Mean-opinion-score (MOS) and pairwise comparison results. Both the MOS and pairwise test results show our proposed method outperforms alternatives, including naive copying and pasting and our method without context. Our method is comparable with complicated TTS, but still has room for improvement. Pairwise scores are given as percent preference for our method.}
\label{tab:listening}
\end{table*}

\subsection{User control}
\label{sec:control}
In addition to automatically adjusting unnatural prosody for text-based speech editing, our method allows users to directly modify the prosody. That is, users of our system can optionally provide explicit pitch and/or duration information to control the prosody of the output speech and refine automatically generated results. If only a subsequence of the full prosody is given (e.g, the pitch values of a word or two or the duration values of a couple phonemes), the remainder is predicted by a neural network to sound natural relative to the user-provided pitch and duration.  In addition, our system can generate a variety of candidate prosodies and allows users to select their favorite. This property is greatly beneficial for our real-world target application and not possible using many existing neural prosody techniques.

\section{Experimental design}

\subsection{Task}
We evaluate our proposed system on the task of copying and pasting a phrase from the text and speech of one sentence to another sentence containing the same phrase. We call this task ``replacement'', and use it to study prosody errors across a variety of alternative methods and compare against a perfect ground truth, unedited recording. We perform replacement on repeated phrases of two-to-five words for each dataset listed below. Given that the text is unchanged, there is a high likelihood of the prosody sounding correct without any additional modification. We manually select 20 examples from each speaker where there is a variety of noticeable prosody errors due to direct waveform editing. This is our ``naive'' condition. All concatenations of waveforms are performed using a 20 millisecond equal-power crossfade to prevent waveform boundary artifacts. We use A-weighted loudness matching \cite{mccurdy1936tentative} to ensure that the volume of the pasted speech is reasonable given the speech context. Our goal is to not only understand how our proposed method can modify the prosody compared to a naive method, but also understand what components of our system are most useful. We use Amazon Mechanical Turk (AMT) to conduct both a mean opinion score (MOS) test and pairwise comparison test against alternative approaches. We additionally use a basic pretest to make sure all subjects have proper equipment and hearing ability.

\subsection{Data}
We demonstrate our system with two single-speaker datasets: one male speaker and one female. Our female speaker is LJSpeech~\cite{ljspeech17}, a subset of books read by Linda Johnson (reader ID 11049 on LibriVox). We preprocess LJSpeech using HiFi-GAN~\cite{su2020hifi} to remove reverberation. For our male speaker, we use the book \textit{The Everyday Life of Abraham Lincoln} read by Bill Boerst (reader ID 4788 on LibriVox). We use the Gentle forced aligner~\cite{ochshorn2016gentle} to break the long-form speech and text into sentences for training. We downsample all speech recordings to a 16 kHz sampling rate.

\subsection{Training}
Our pitch generator~\cite{morrison2020controllable} and denoising~\cite{su2020hifi} networks are pretrained according to their original implementations. Our duration generator is trained in a single-speaker fashion on each of our datasets. We use two hours of speech for training and two hours for validation. We train the duration generator for 30 epochs with a batch size of 64. Training the duration generator takes less than 30 minutes on one Tesla V100 GPU.

During training, we randomly provide short sequences of the ground truth phoneme duration as input features. We randomly select half of the training samples to have $k$ adjacent ground truth durations, where $k \sim \text{Uniform}(0, 24)$, so that the model learns to fill in remaining durations in a context-aware manner. During inference, the phoneme durations of the speech context are similarly provided as input features. The remaining, unspecified durations are predicted based on the durations of the context, which encourages prosodic continuity. We call this method \textit{subsequence prediction}, and it is also used by our pitch generator to enable similar control over the generated pitch~\cite{morrison2020controllable}. We use subsequence prediction to condition the pitch and duration generators on the proceeding and following speech prosody. This makes our prosody generators context-aware.

\subsection{Method comparisons}

We include several additional test conditions in our listening study including listening test anchors, ablations, and baselines. We use the original speech as a high-quality anchor (Original). For a low-quality anchor, we use an average condition (Average), which uses monotone pitch with constant pitch at the mean frequency of the speech context and per-phoneme duration averages computed on the training data. We ablate our duration generator by using the durations of the baseline condition instead of generating new durations \mbox{(-Duration)}. We similarly ablate our pitch generator by using the pitch of the naive cut condition \mbox{(-Pitch)}. We ablate HiFi-GAN by using the raw TD-PSOLA output without postprocessing \mbox{(-HiFi-GAN)}. We examine the role of context-awareness by producing examples that use our proposed method without subsequence prediction \mbox{(-Context)}. We compare our method to a baseline that uses prosody generated by \mbox{a multispeaker model based on Tacotron 2}~\cite{jia2018transfer} and vocoded with WaveGlow~\cite{prenger2019waveglow} using the full text of each example. We extract the prosody using pitch-tracking and phoneme alignment, and apply the prosody via TD-PSOLA. This allows us to directly compare the prosody generated by the Tacotron-based model to our methods without the additional comparison of TD-PSOLA and WaveGlow.


\section{Results}
\label{sec:resul}
We report our listening test results for both the MOS and pairwise comparison tests in~\tref{tab:listening}. A total of 159 and 220 unique listeners participated for the MOS and pairwise comparison tasks, respectively.  We obtain a total of 16 answers per voice per condition per utterance.

First, we compare our proposed system with and without context-awareness \mbox{(-Context)}. We see that context improves MOS results by a noticeable amount ($0.41$) as well as pairwise preference ratings. This result demonstrates the significance of context-awareness in deep learning models that perform speech editing tasks.

Second, we compare our method to a state-of-the-art TTS method (Tacotron). When we look at both the MOS and pairwise comparison scores, we find our method slightly outperforms Tacotron for the task of text-based speech editing. Given the relative simplicity of our prosody generation networks, this is remarkable, as it shows that simple prosody generators can achieve comparable prosody naturalness without training a complex, multi-speaker acoustic feature predictor or neural vocoder.

%

Third, we compare our method to our own Naive cut-copy-paste implementation. Both MOS and pairwise comparison results show our proposed method performs better. While it might seem like this should be easy to outperform, it is important to note that the Naive method uses only naturally recorded audio and has no time-stretching, pitch-shifting, or other vocoder artifacts by construction. The manual selection process for this condition produces examples that requires large pitch-shifting and time-stretching operations. TD-PSOLA creates larger artifacts during larger manipulations. Our method surpasses this Naive baseline despite inducing a worst-case scenario for the naturalness of TD-PSOLA.

Fourth, we ablate each of our pitch (-Pitch) and duration (-Duration) generators. We see that each of these components is important, but the effect of pitch manipulation is more significant. This result is inline with to our initial analysis of phoneme duration being unimodal and suggests that further work on context-aware pitch generation may be beneficial.

Finally, we compare our method with and without HiFi-GAN. Both MOS and pairwise comparison results suggest that HiFi-GAN mitigates speech manipulation artifacts caused by time-stretching and pitch-shifting. We suspect that further advancements in neural denoising and dereverberation systems will directly benefit our proposed method.

\section{Conclusion}
We propose a system that enables high-fidelity context-aware text-based speech editing. Our system trains quickly and automatically infers a natural sounding prosody relative to the speech surrounding the edit region, but also allows explicit user adjustment for complete creative control. We anticipate that these additional controls for the production and editing of speech can expedite the creative workflow of both novice and expert content creators. Toward that end, our future work could explore using an efficient, high-fidelity neural vocoder instead of TD-PSOLA, or replacing HiFi-GAN with a denoiser catered specifically for TD-PSOLA that can correct formant errors. The latter would be especially useful as TD-PSOLA can perform real-time manipulation even at high sampling rates.

\bibliographystyle{IEEEbib}
\bibliography{refs}

\end{document}